\documentclass[runningheads,a4paper]{llncs}

\usepackage{amssymb,amsmath}
\usepackage{graphicx}
\usepackage[tight,footnotesize]{subfigure}
\usepackage[font=footnotesize]{subfig}
\usepackage{url}
\usepackage{footnote}
\usepackage{tabularx}
\usepackage{booktabs}

\urldef{\mailsa}\path|temerina@frias.uni-freiburg.de|
   
\newcommand{\keywords}[1]{\par\addvspace\baselineskip
\noindent\keywordname\enspace\ignorespaces#1}

\begin{document}

\mainmatter  

\title{Deciding when to stop: Efficient stopping of active learning guided drug-target prediction}

\titlerunning{Deciding when to stop}

%
%
\author{Maja Temerinac-Ott$^{1}$%
\thanks{This study was supported by BMBF e:BIO grant \textit{Microsystems}, FKZ0316185. This paper was selected for oral presentation at RECOMB 2015 and an abstract is published in the conference proceedings.}%
\and Armaghan W. Naik$^{2}$ \and Robert F. Murphy$^{1,2,3}$}
\authorrunning{Temerinac-Ott, M., Naik, A.W., Murphy, R.F.}

\institute{$^{1}$ Freiburg Institute for Advanced Studies, University of Freiburg, Germany \\
     $^{2}$ Computational Biology Department, Carnegie Mellon University, Pittsburgh, PA, USA\\
$^{3}$ Departments of Biological Sciences, Biomedical Engineering and Machine Learning, Carnegie Mellon
University, Pittsburgh, PA, USA\\
\mailsa\\
}

\toctitle{RECOMB 2015}
\tocauthor{Deciding when to stop: Efficient stopping of active learning guided drug-target prediction}
\maketitle

\begin{abstract}
Active learning has shown to reduce the number of experiments needed to obtain high-confidence drug-target predictions. However, in order to actually save experiments using active learning, it is crucial to have a method to evaluate the quality of the current prediction and decide when to stop the experimentation process. Only by applying reliable stoping criteria to active learning, time and costs in the experimental process can be actually saved. We compute active learning traces on simulated drug-target matrices in order to learn a regression model for the accuracy of the active learner. By analyzing the performance of the regression model on simulated data, we design stopping criteria for previously unseen experimental matrices. We demonstrate on four previously characterized drug effect data sets that applying the stopping criteria can result in upto $40\%$ savings of the total experiments for highly accurate predictions.

\end{abstract}

\section{Introduction}
\label{sec:intro}
A critical step in developing new therapeutics is frequently to conduct large scale searches for potential drugs that can affect a desired target.  Recently, it has become clear that finding successful drugs also requires searching for the absence of undesired effects on other targets.  This need cannot be met by exhaustive experimentation, but selective experimentation driven by machine learning (a process referred to as \textit{active learning}) may provide an alternative  (\cite{Murphy2011}).  The heart of active learning is having good predictive models to guide experimentation. Recent studies show that drug-target prediction algorithms can speed-up the discovery of new drugs (e.g., \cite{Besnard2012,Paolini2006,Reymon2010,Keiser2009}).

Current drug-target prediction methods are coarse grained over at most a handful of 'campaigns'. In these, a classifier is trained with relatively large amounts of training data resulting from exhaustive screening, and then verified on a small test set. These data are generally identified manually, and limited to human 'expert' knowledge. This process is generally only performed once, or at most a handful of times due to the expense of exhaustive screening over many compounds. This procedure limits the generalization capability of the model and does not allow for an optimal exploration of the drug-target interaction space. Alternatively, active learning methods can be used to iteratively build a model of drug-target interactions. Instead of relying on large training data sets, the active learning procedure enlarges the training set stepwise, guided by the predictions on small, automatically-selected test sets. Thus time and costs are spent on improving the general model rather than having the verification of a small specific model that does not account for the large space of chemical compounds. The general model has the potential to predict side-effects early on in the drug design process, since a larger number of drugs are considered in the drug-target prediction matrix. A critical point when using active learning to guide experimentation is to decide when to stop, since the goal is to perform as few as possible experiments in order to have the best model. The best stopping time is reached when adding new experiments to the training set will not improve the accuracy on the test set. The difficulty, of course, is that calculating the true accuracy of the model requires all of the data. Therfore, reliable methods for predicting the accuracy of the current model during an active learning cycle are desired. Due to experimental cost and time restrictions, the best stopping time might not be desirable, so it would be helpful to stop earlier when a predifined confidence on the output of the model is reached. 

Previous work in this area has generally addressed active learning methods or drug-target prediction methods, but rerely both.  For example, active learning has been used to identify active compounds from a large pool of compounds targeting a single molecule \cite{Warmuth2003}. Active learning has also been applied in the context of cancer research  \cite{Danziger2007}. Several methods for drug-target prediction without active learning have been proposed recently \cite{Yamanishi2010,Atias2011,Campillos2008,Alaimo2013,Cheng2012,DBLP:conf/kdd/ZhengDMZ13,Bleakley2009} and remain an active area of research. The focus of this work is not to promote a particualar drug-target prediction method, but to show using matrix factorization as an example how drug-target prediction can be combined with active learning and lead to reductions of experimentation cost. Initial results on applying active learning for drug-target prediction on multiple drugs and multiple targets simultaneously have been reported \cite{Naik2013,Kangas2014}, with and without requiring prior knowledge of drug or target similarities. In \cite{Kangas2014} the benefits of active learning on a large dataset from PubChem are reported, however without applying the stopping rule. In \cite{Naik2013} an intital method for predicting the accuracy of active learning traces is presented, however it was not applied to the particular problem of drug-target pediction.

Several stopping rules for active learning have been considered in the past \cite{Laws2008,Vlachos2008,Zhu2010}, however there has been little analysis of which performs the best in general. Four simple stopping criteria based on confidence estimation over the unlabeled data pool and the label consistency between neighboring training rounds of active learning have been presented \cite{Zhu2010}. Instead of using a single criterion to stop, combining different stopping criteria in a feature vector describing the active learning trajectory has been proposed in \cite{Naik2013}. The features of trajectories on simulated data are used to train a regression function in order to predict the accuracy of active learning algorithms on unseen simulated data. We will follow this approach and adopt it to the binary drug-target prediction case. 

The major goals of our active learning system are: (1) We want to have a fast and reliable method to elucidate drug-target interactions. (2) Previous knowledge on similarities between drugs and similarities between targets should be included in the model, so that predictions for new drugs or targets (for which no experiments are available) are possible. (3) The number of experiments required to make confident predictions should be systematically reduced. (4) An efficient stopping rule for ending the active learning process should be designed.

Previously, kernel-based matrix factorization (\cite{Giannakis2013}) has  been shown to provide good models of drug-target interactions (\cite{Goenen2012}). In the kernelized Bayesian matrix factorization (KBMF) algorithm (\cite{Goenen2012,Goenen2013}), the drug-target interaction matrix is factorized by projecting the drugs and the targets into a common subspace, where the projected drug matrix and the projected target matrix can be multiplied in order to produce a prediction for the drug-target interaction matrix. The entries of the prediction matrix are modeled using truncated normal distributions.  The projected drug matrix and target matrix are factored using two different kernels: a drug specific kernel and a target specific kernel. A kernel encodes the similarity between the drug and the target features. Thus prior information can be easily inserted to the model. Furthermore, the knowledge of the full interaction matrix is not needed in order to make predictions for new drugs, which is not the case for previous methods (i.e. \cite{Atias2011}).

The main contributions of this work are: (i)	We use KBMF to construct a powerful and practical active learning strategy for analyzing drug-target interactions.  (ii) We extend  previous work \cite{Naik2013} on estimating the accuracy of active learning predictions to the KBMF case and show how it can be used to construct a stopping rule for experimentation. (iii) We provide a proof of concept through evaluation of the method on four data sets previously used for modeling of drug-target interactions  \cite{Yamanishi2008}. (iv) We show the superiority of the proposed active learning approach compared to random choice of an equivalent number of experiments.

\section{Methods}
\subsection*{Active Learning Framework}
\begin{figure*}[!t] 
   \begin{center}
  \includegraphics[height=6 cm]{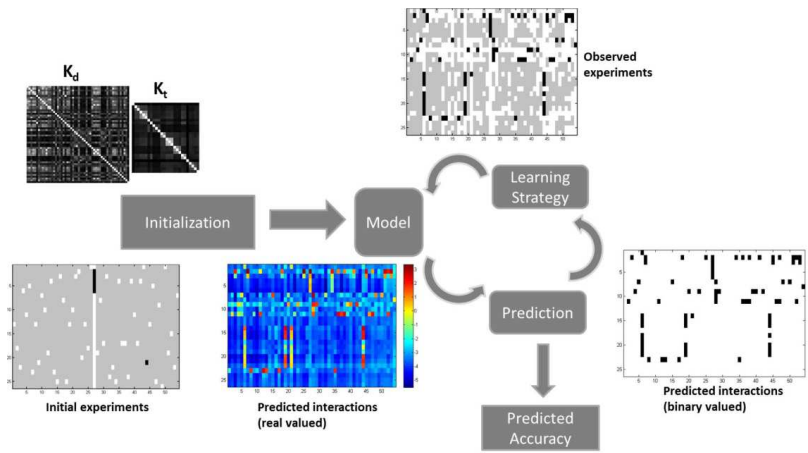}
   \end{center}
   \caption{The major components of an active learning framework. The entries of the matrix are color coded: label not known (light gray), interaction (black), no interaction (white). At initialization a subset of known labels for the interactions matrix and the drug and target kernels $\mathbf{K_d}$ and $\mathbf{K_t}$ are provided. In each round of the active learning algorithm, the labels of the entire interaction matrix are predicted and used to determine which labels to query next. In this figure, the dark red values represent a high probability for a hit, whereas the dark blue values represent a high probability for a miss.}

	\label{fig:AL_anatomy}
\end{figure*}

An active learning method is an iterative process composed of four components: the initialization, the model, the active learning strategy and an accuracy measure for the predicted output in each step (Fig. \ref{fig:AL_anatomy}). Most active learning papers focus on the second and third components. The active learning framework starts with an initialization strategy which is followed by the generation of a model. The model is used to make predictions, in our application drug-target interactions are predicted. Interactions can be measured by performing an \textit{experiment}, i.e. a direct assay of drug-target interaction (e.g., in cell extracts). Based on the predictions, an active learning strategy is applied to query new experiments (labels) which will improve the model. We use batchwise learning, where a fixed number of experiments is queried in each training round. Each training round defines a \textit{time-point} in the active learning process and is measured by the number of batches of experiments performed. For each time-point the accuracy of the model is predicted. The process is stopped for example, if a certain budget for performing experiments is reached or the predicted accuracy of the model is high enough. We assume equal cost for each experiment.

\subsection*{Data Representation}
We use interaction matrices $\mathbf Y \in \{-1,1\}^{N \times M}$ to represent drug-target interactions. We assume that the outcome of the experiment determines the ground truth label $l \in \mathcal L = \{-1,1\}$ for an interaction matrix entry. $N \in \mathbb N$ is the number of drugs, $M \in \mathbb N$ is the number of targets. Knowledge of the interaction between a drug $d \in \{1,2,...,N\}$ and a target $t \in \{1,2,...,M\} $ is ternary encoded in the \textit{experimental matrix} $\mathbf X$: $+1$ for an interaction, $-1$ for lack of interaction, and $0$ to denote experiments which have not yet been performed. Hereby, the set of remaining experiments (unlabeled data) will be denoted by $\mathcal{X} = \{x = (d,t)|\mathbf X(x) = 0  \}$. Therefore, we consider a semi-supervised binary labeling problem where the sign of the label indicates the interaction status between a drug and a target.

\subsection*{Kernelized Bayesian Matrix Factorization (KBMF)}
\label{sec:KBMF}
As described previously \cite{Goenen2012,Goenen2013}, KBMF can be effectively applied to model drug-target interactions.  It approximates the interaction matrix by projecting the drug kernel $\mathbf{K_d}\in \mathbb{R}^{N \times N}$ and the target kernel $\mathbf{K_t}\in \mathbb{R}^{M \times M}$ into a common subspace of dimension $R \in \mathbb N$ such that the interaction matrix $\mathbf{Y}$ can be reconstructed from the sign of its prediction matrix $\mathbf{F} \in \mathbb{R}^{M \times N}$:

\begin{equation}
	{\mathbf{\hat Y}}(d,t) = \left\{ 
  \begin{array}{r l}
    1 & \quad \text{if $\mathbf{F}(d,t)>0$}\\
    -1 & \quad \text{else}.
  \end{array} \right. 
\end{equation}

The prediction matrix $\mathbf{F}$ is a product of the projected kernel matrices:
\begin{equation}
	\mathbf{F} = ((\mathbf{A_d})^T\mathbf{K_{d}})^T((\mathbf{A_t})^T\mathbf{K_{t}}), 
\end{equation}
where $\mathbf{A_d}\in \mathbb{R}^{N \times R}$ and $\mathbf{A_t}\in \mathbb{R}^{M \times R}$ are subspace transformation matrices computed by the variational Bayes algorithm  \cite{Goenen2012,Goenen2013} using the values of the experimental matrix $\mathbf X$. The dimension $R$ of the subspace is a free parameter; we used the value of 20 previously determined to be optimal for these datasets \cite{Goenen2013}. The entries of the kernel matrix $\mathbf{K_d}$ and $\mathbf{K_t}$ are a measure of the pairwise similarities between drugs and targets respectively. The similarity matrices provided by Yamanishi et al. \cite{Yamanishi2008} and the KBMF implementation of semi-supervised classification \footnote{\url{http://research.ics.aalto.fi/mi/software/kbmf/}} provided by Goenen \cite{Goenen2013} were used.

\subsection*{Initialization and experiment selection}\label{sec:BatchSelection}
Our initialization strategy is to select a random column and one random experiment from each row of the experimental matrix $\mathbf X$.

\subsubsection*{Uncertainty sampling}
We use uncertainty sampling (\cite{Lewis1994}) to form a batch of experiments $\{x_1,..,x_K\}\in \mathcal X$ by greedily choosing the $K\in \mathbb N$ experiments with the greatest uncertainty function $U$ (\cite{Zhu2010}):

\begin{equation}
 U(x)=-\sum_{l \in \mathcal L} P(l|x)log P(l|x).
\end{equation}

For the KBMF case the posterior probability is computed by the sigmoid function from the predicted interactions: 
\begin{equation}
	P(l=1|x) = \frac{1}{1+\exp(-\mathbf F(x))},%
\end{equation}
and $P(l=-1|x) = 1-P(l=1|x)$ for no interaction respectively.
\subsection*{Stopping Rule}\label{sec:Confidence}

In order to stop the active learning process, a method is needed to predict the accuracy of the model for a given time-point along with the confidence of that prediction. As proposed previously in \cite{Naik2013}, the accuracy of a model at a given point in an active learning process can be predicted using a regression function trained for other, similar experimental spaces. The fully observed drug-target space is characterized by two measures, uniqueness ($u$) and responsiveness ($r$) \cite{Naik2013} defined by:

\begin{eqnarray}
r &=& \frac{1}{N\cdot M}\sum_{d,t,Y(d,t)=1}\mathbf Y(d,t)\\
u &=& \frac{uRows(\mathbf Y)+ uColumns(\mathbf Y)}{N+M},
\end{eqnarray}

where $uRows(.)$ and $uColumns$ compute the number of unique rows and unique columns of a matrix.

The uniqueness and responsiveness are values in the range $[0,1]$ and characterize the interaction matrix. Responsiveness measures the percentage of interactions in the matrix. Uniqueness is a measure of independence of the rows and columns in the matrix. The higher the value for uniqueness is, the more difficult it is to make predictions.

These two measures have two purposes: (1) They are used to compute features for a time-step in our current active learning process. (2) They can be used to generate simulation data having similar properties to the measured experimental data. 

Each time-point $t_i$ is described by a vector of 13 features $f_{t_i} \in \mathbb R^{13}$, defined as:
\begin{itemize}
 \item $f(1), f(2)$: average observed responsiveness across columns (respectively rows)
 \item $f(3), f(4)$: average predicted responsiveness across columns (respectively rows) 	  
\item $f(5)$: average difference in predictions from last prediction for current time-point ($t_{i}$) 
\item $f(6)$: average difference in predictions from last prediction for previous time-point ($t_{i-1}$)
\item $f(7)$: fraction of predictions at $t_{i-1}$ observed as responsive ($l=1$) at $t_{i}$
\item $f(8),..,f(10)$: minimum, maximum and mean number of experiments that have been performed for any drug
\item $f(11),..,f(13)$: minimum, maximum and mean number of experiments that have been performed for any target
\end{itemize}

These features are normalized to the range $[0..1]$ and are further extended by computing the square root of their pairwise products. 

To learn the accuracy predictor via simulation data, interaction matrices of size $50 \times 50$ were randomly sampled in the grid of uniqueness and responsiveness parameters $5\%,10\%,\ldots,95\%$. For each interaction matrix we derived 'perfect' Gaussian similarity kernels $\mathbf{K_d}, \mathbf{K_t}$ by pairwise distances of the column-space and row-space, respectively. These were disrupted by forcing $0\%,5\%,10\%$ of the kernel entries to the value 1 and regularized to ensure positive semidefiniteness. Features computed from trajectories of the uncertainty sampling active learner on these data were collected; for each trajectory we also measured the accuracy of prediction against the ground truth. A linear model of these features against adjusted accuracies (accuracy above the fraction of experiments performed so far) was fitted by lasso regression (\cite{Tibshirani1996}). The lasso regularization parameter was chosen by 11-fold cross validation under squared loss, with holdout granularity at the level of trajectories. To make accuracy predictions from adjusted accuracy predictions, we added the fraction of experiments performed so far.

\section{Results}

For validation of our method, experiments are performed on four data sets extracted from the KEGG BRITE (\cite{Kanehisa2006}), BRENDA \cite{Schomburg2004} , SuperTarget \cite{Guenther2008} and DrugBank \cite{Wishart2008} databases, previously described by Yamanishi et al \cite{Yamanishi2008} \footnote{\url{http://web.kuicr.kyoto-u.ac.jp/supp/yoshi/drugtarget/}}. The data set consists of four drug-target interaction matrices: Nuclear Receptor, GPCR, Ion Channel and Enzyme.

\subsection*{Comparison of active and random learning strategies}
\begin{figure*}[!t]
   \begin{center}
		\subfigure[Nuclear Receptor]{\includegraphics[height=3.8 cm]{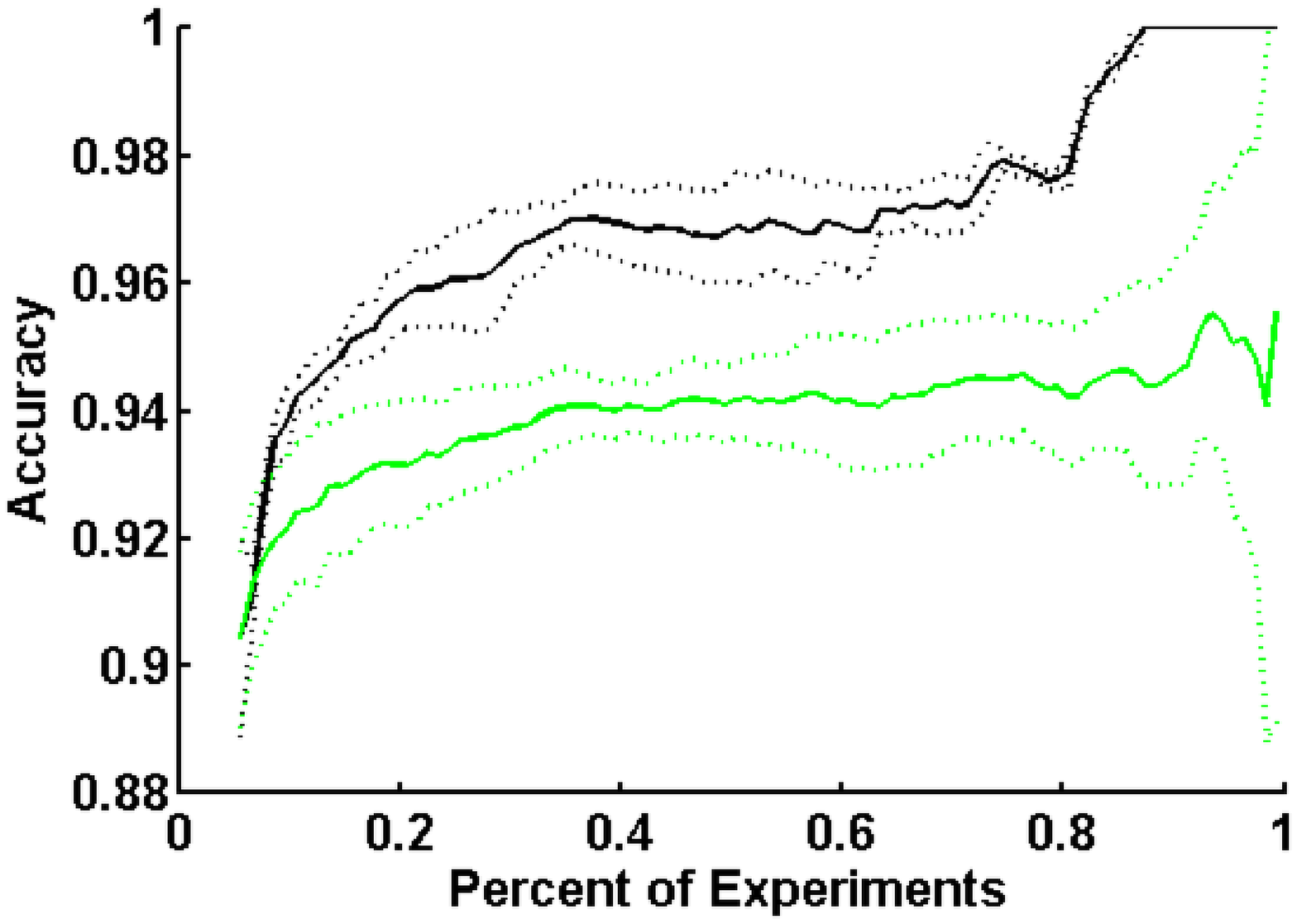}} \hspace{0.3cm}
   \subfigure[GPCR]{ \includegraphics[height=3.8 cm]{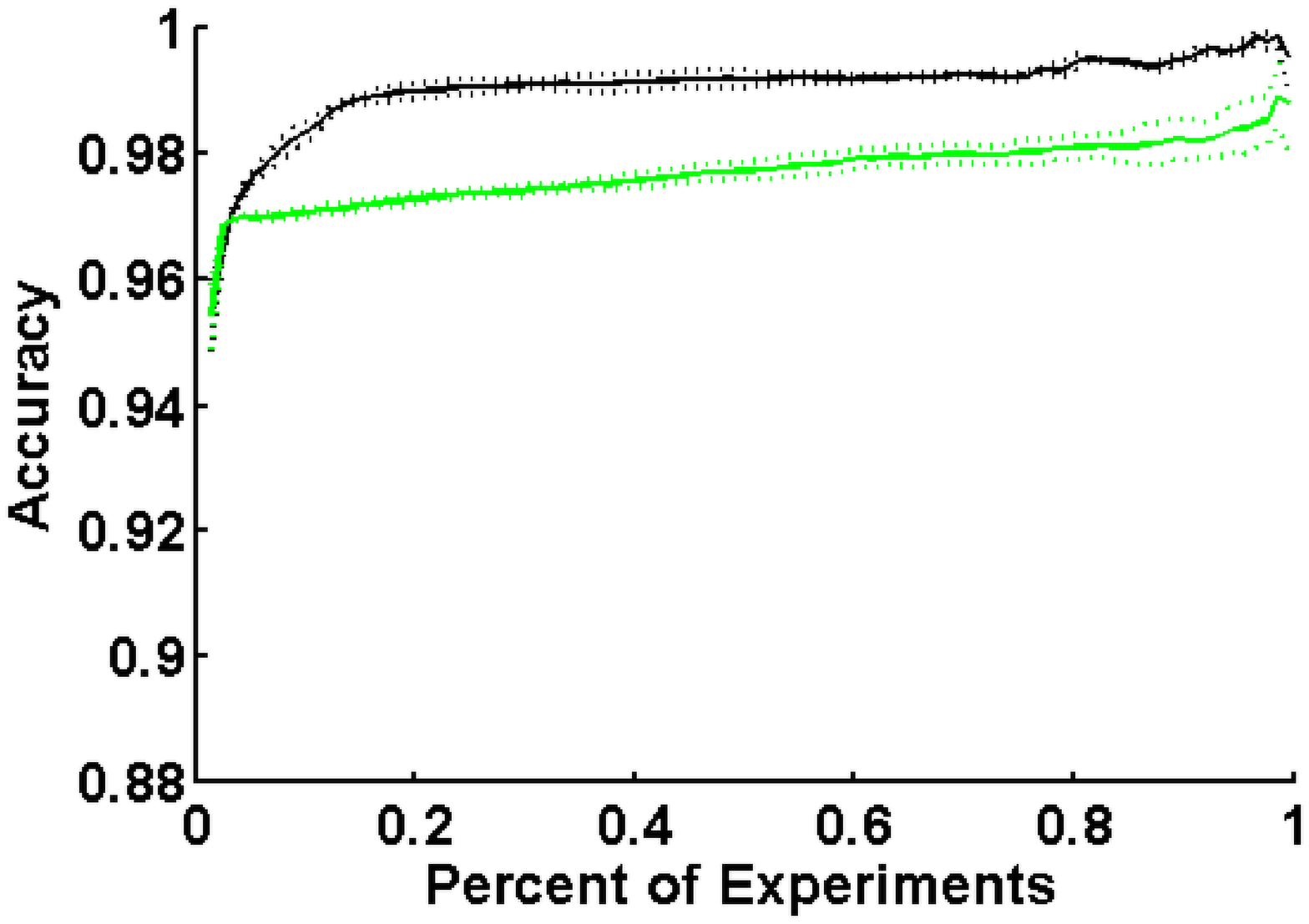}}\\
   \subfigure[Ion Channel]{ \includegraphics[height=3.8 cm]{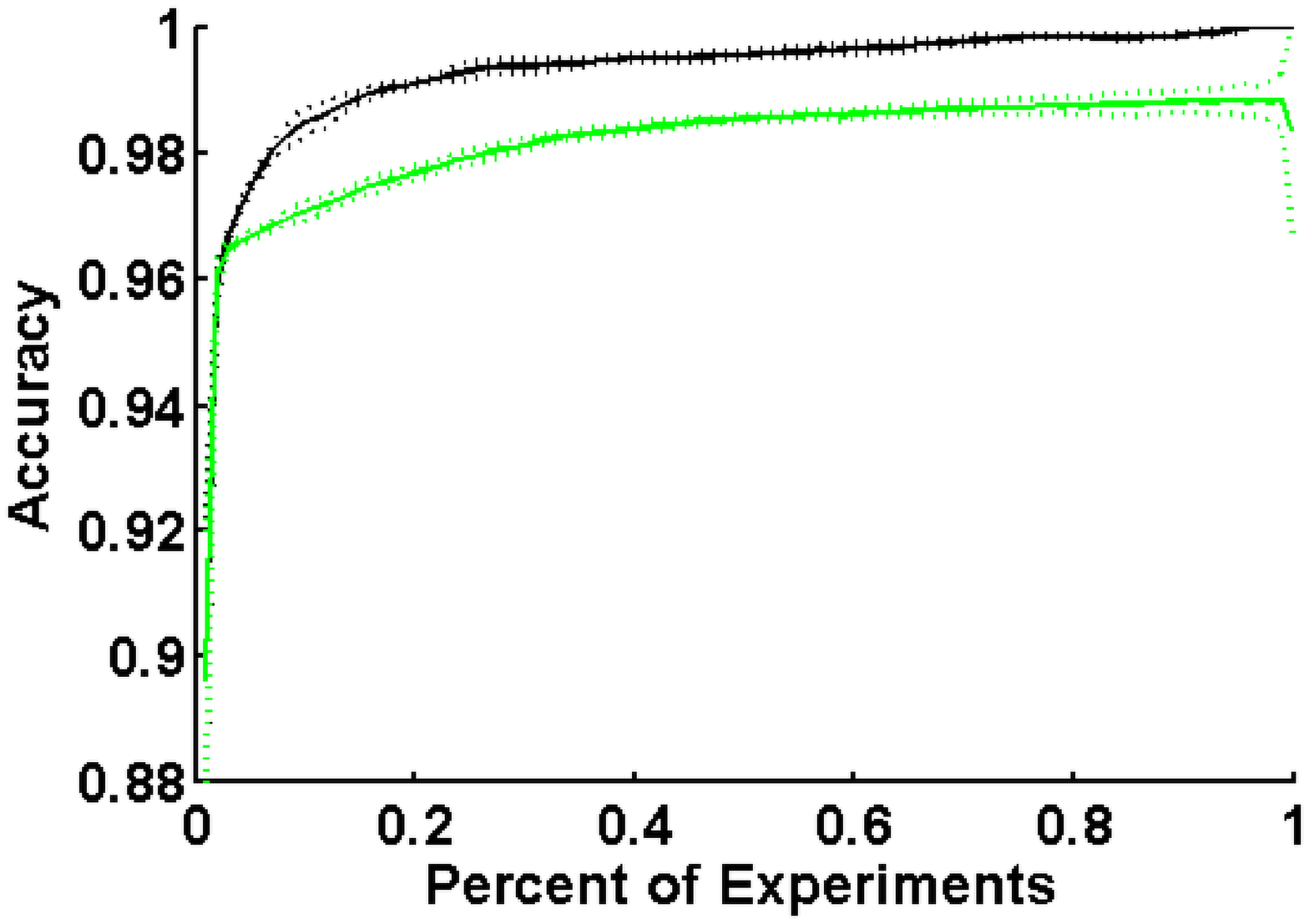}}\hspace{0.3cm}
	\subfigure[Enzyme]{ \includegraphics[height=3.8 cm]{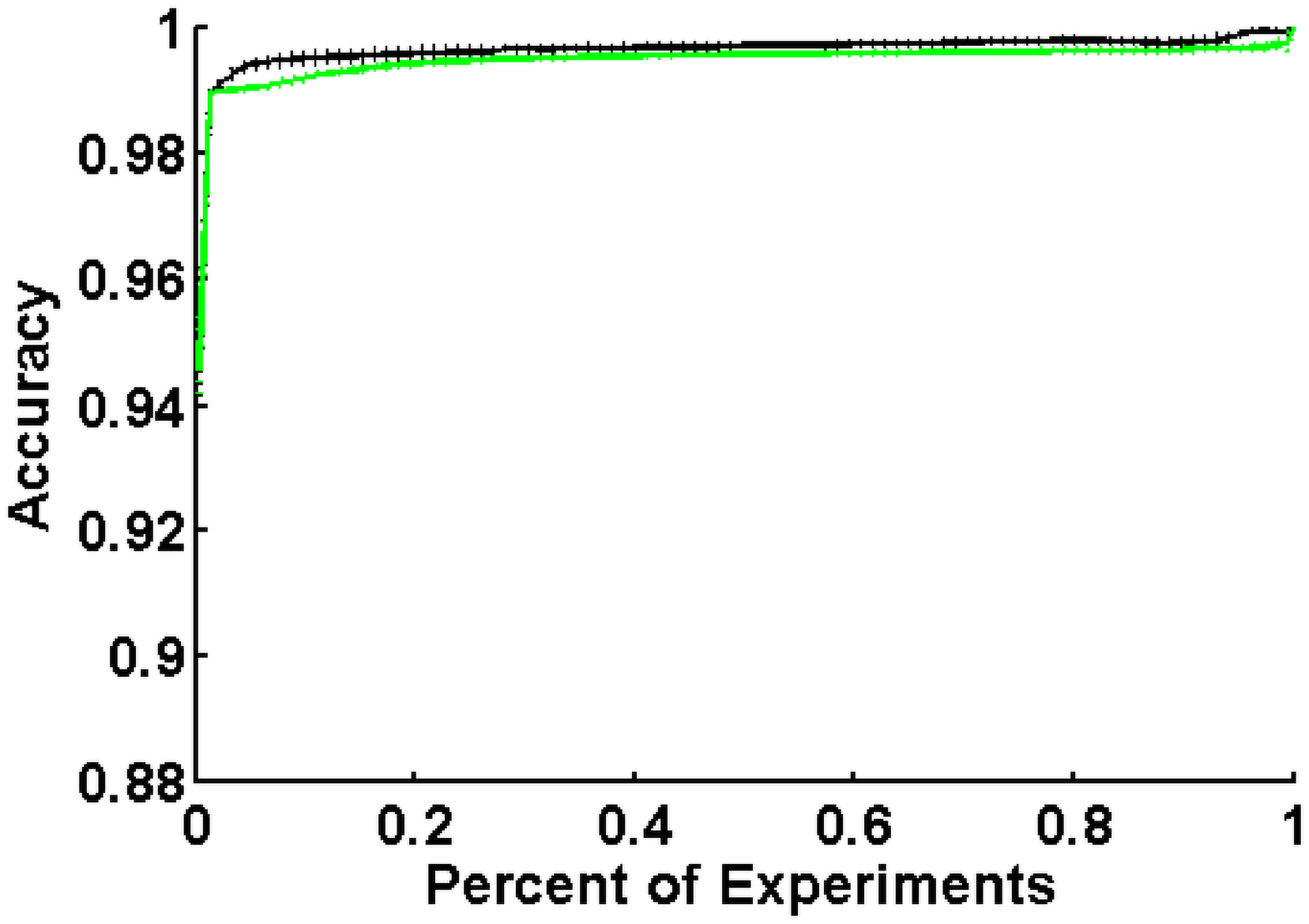}}
   \end{center}
   \caption{Comparison of random sampling (green) to uncertainty sampling (black) on the four data sets. The solid and the dotted line represent respectively the mean and the standard deviation of 5 random initializations. For random sampling five random runs were performed for each initialization.}\label{fig:ComparisonRandom} 
\end{figure*}

In order to evaluate the efficiency of active learning methods, we compared the uncertainty sampling strategy  with random choice of experiments in each time-step. On all four data sets the active learning strategy outperformed the random strategy (Fig. \ref{fig:ComparisonRandom}). On the GPCR and the Ion Channel dataset the active learning strategy using uncertainty sampling reaches $99\%$ accuracy 5-6 times faster than the random strategy.

\subsection*{Predicting the accuracy of the model }
\begin{figure*}[!t]
   \begin{center}
	\subfigure[Nuclear Receptor]{\includegraphics[height=3.8 cm]{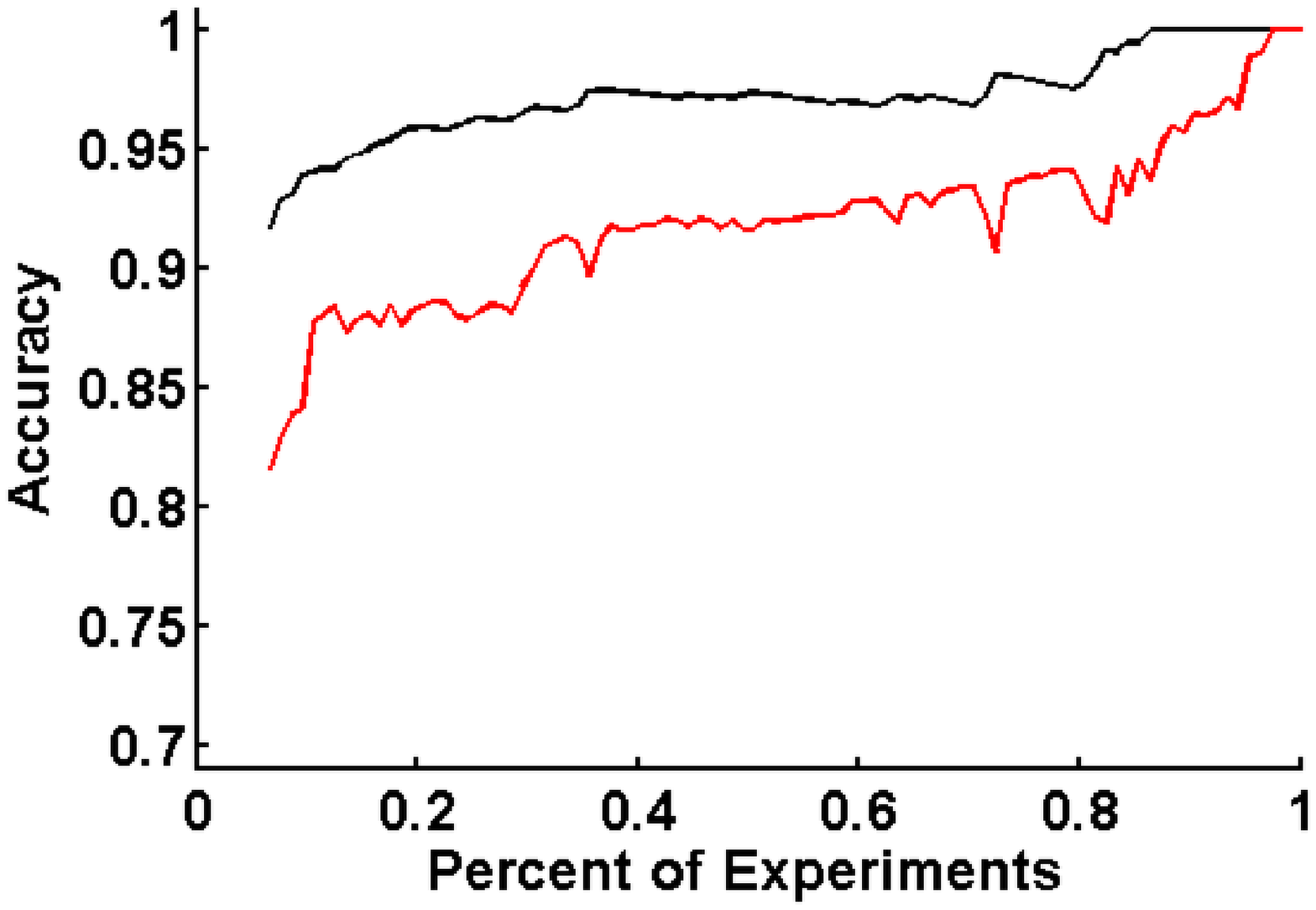}} \hspace{0.3cm}
   \subfigure[GPCR]{ \includegraphics[height=3.8 cm]{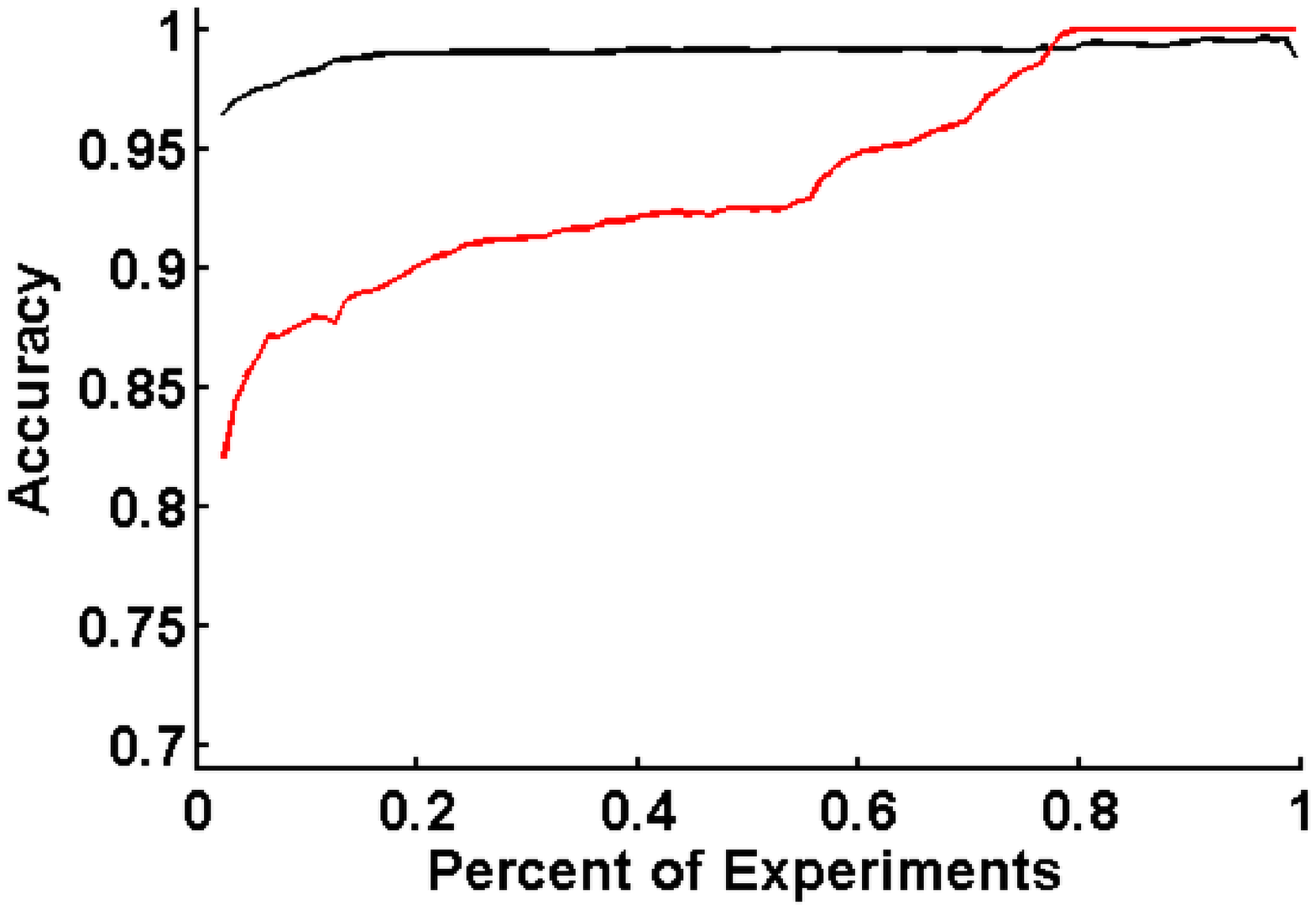}}\\
   \subfigure[Ion Channel]{ \includegraphics[height=3.8 cm]{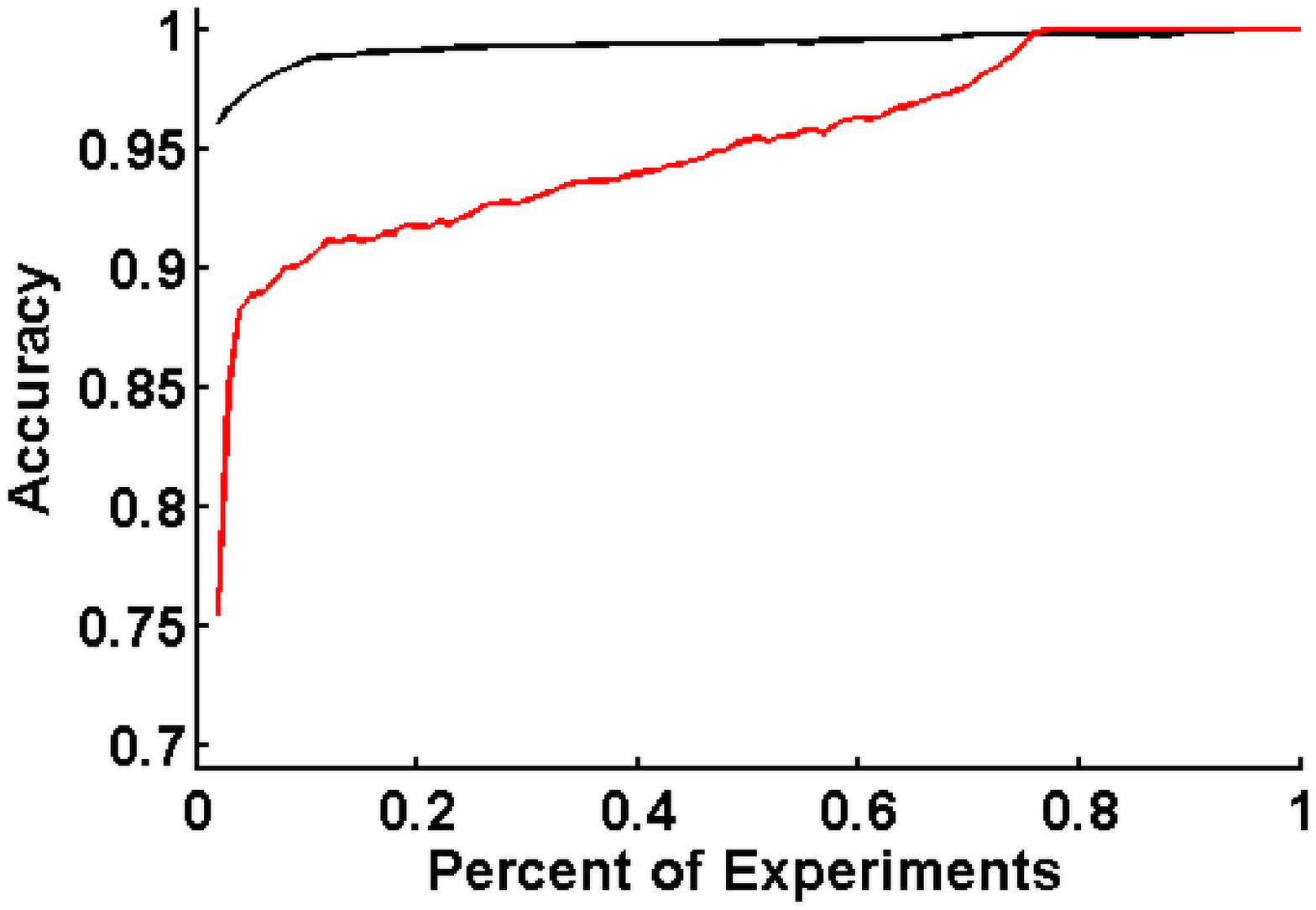}}\hspace{0.3cm}
	\subfigure[Enzyme]{ \includegraphics[height=3.8 cm]{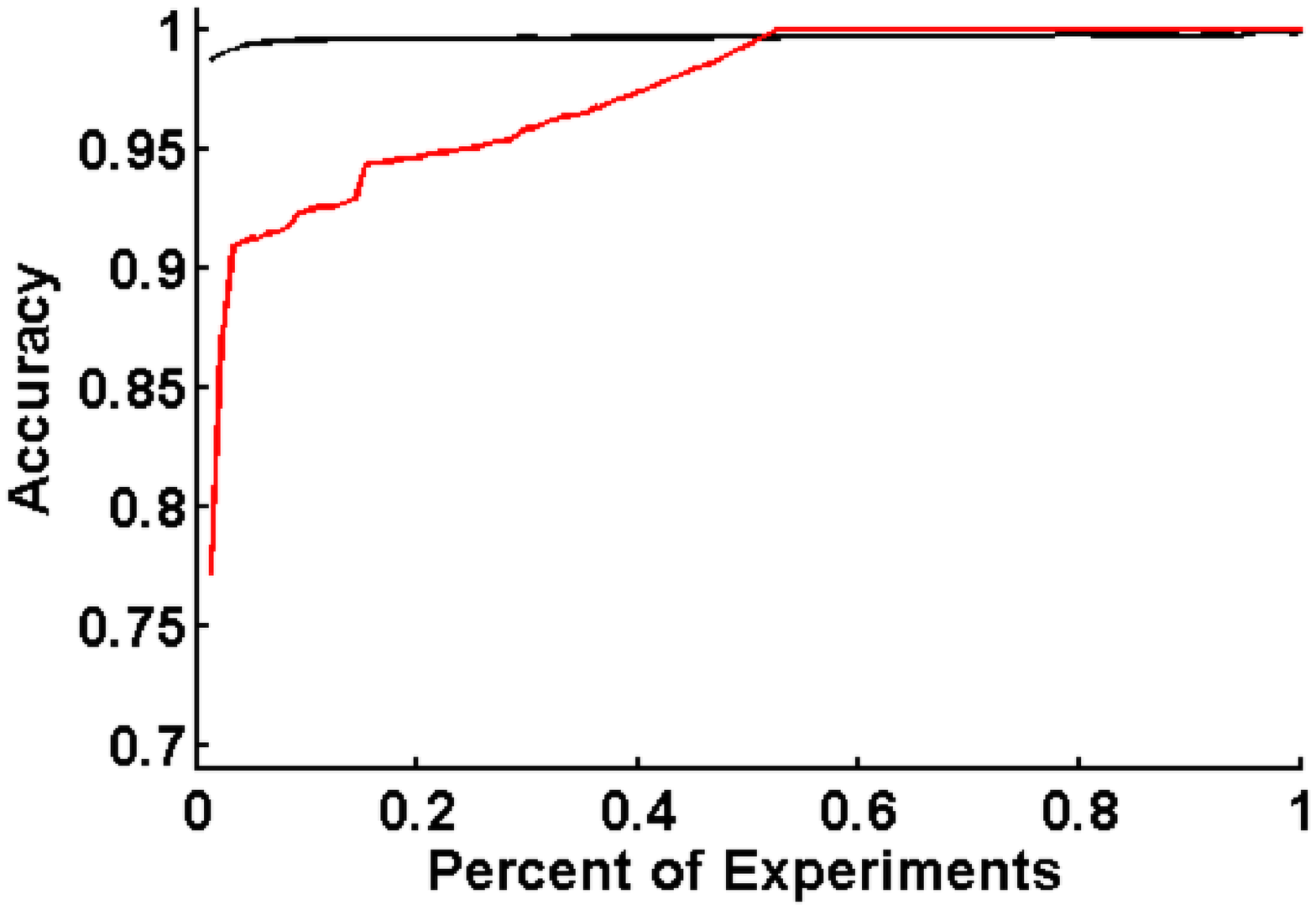}}
   \end{center}
   \caption{The true accuracy (black) and the predicted accuracy (red) are shown for the four data sets.}\label{fig:Regression} 
\end{figure*}

As discussed in the introduction, in practice we require a mechanism to decide when to stop experimentation. It is not enough to have a good active learning method without the possibility to evaluate the accuracy of the whole model apart from acquiring all the data. Therefore we have simulated interaction matrices with uniqueness, responsiveness in the range $[0.05..0.95]$ and kernel noise in the range $[0..0.1]$. We then performed active learning simulations using our KBMF model and uncertainty sampling and learned a regression function for the predicted accuracy. The results of applying the regression function to the computed features at each time point are shown in red in Fig. \ref{fig:Regression} for the four data sets. On all four data sets, the predicted accuracy of $90\%$ guarantees the true accuracy to be at least $90\%$, and the predicted accuracies are a reasonable lower estimate for the true accuracy. 

\subsection*{Learning the stopping rule}
\begin{figure*}[!t]
   \begin{center}
	 \includegraphics[height=4.3 cm]{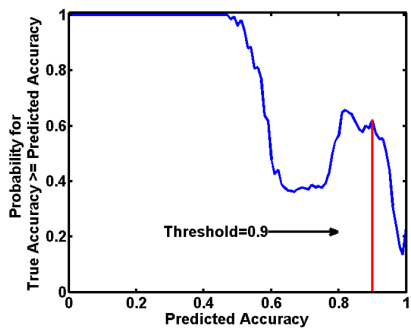}
   \end{center}
   \caption{The probability that a predicted accuracy is below or equal to the true accuracy is plotted against the threshold.}\label{fig:PredictionStatistics} 
\end{figure*}

Statistics on the performance of the accuracy predictor in simulations can be used to design a stopping rule \cite{Naik2013}. We adopt this method to determine a threshold for stopping the active learning procedure. The simulated data is used to assess the probability that the true accuracy is greater than or equal to the predicted accuracy using 11-fold cross-validation.  We count for each predicted accuracy value how often the condition was fulfilled and divide it by the total occurrence of this predicted value (Fig. \ref{fig:PredictionStatistics}). As expected, a low predicted accuracy with a value below 0.5 will have a high probability to be measured higher in the actual experiment. Predicted values below 0.5 are not of interest, since the predicted value is too low. In the beginning of the active learning procedure a small amount of data is available, so it is hard to make good predictions about the accuracy of the method. However, the more data is gathered in the active learning procedure, the more confident the predictor gets, reaching a peak for predicting the accuracy of 0.8 and higher for $65\%$ of the cases. For very high accuracies ($>0.95$), the chance that the actual accuracy exceeds the prediction naturally drops drastically. From Fig. \ref{fig:PredictionStatistics} the best threshold to stop lies in the range 0.8 to 0.9. Since higher accuracy values are more desirable, our stopping rule was to terminate the active learning procedure, when the predicted accuracy is 0.9. 

\subsection*{Applying the stopping rule}
In the original KBMF paper (\cite{Goenen2012}), the KBMF classifier was evaluated by 5-fold five cross validation using $80\%$ of drugs for training and $20\%$ of drugs for testing. We wanted to test if a matching accuracy on the test set could be reached by choosing actively a reduced number of experiments for training. In other words, assuming that we get to perform selected experiments drawn from a given set, use them to train a model, and make predictions for a held out set (for which experiments are not possible), can we get an accurate model without doing all experiments?  For this purpose our active learning strategy was modified. We use $1\%$ of drugs as the batch size and select in each run the drugs which the classifier is the most uncertain about. For uncertainty sampling using the predictions of the KBMF classifier, this means that drugs with the maximal mean uncertainty across targets are selected.
 
Instead of using all $80\%$ of drugs for training (\cite{Goenen2012}), we use the predicted accuracy on the training data to stop acquisition. When the predicted accuracy on the training set reaches a threshold, the active learning process is stopped and the AUC value on the test set (the $20\%$ of the drugs which were held out) is reported. The average results after 5-fold five cross validation are reported in Table \ref{tab:AUC_comparison_holdout}. By using the stopping-rule on all four data sets, only half of the drugs were needed for training to reach a similar AUC value to that when using all 80\% of the drugs for training.

\begin{table}[!t]
\caption{Average AUC on hold out data and percentage of experiments after applying our stopping rule. The average AUC obtained on held out data using $80\%$ of the data for training \cite{Goenen2012} is compared with the average AUC obtained by training with only the listed percentage of experiments obtained by applying the stopping rule.  The percentage of experiments can be halved by using the proposed stopping rule.  \label{tab:AUC_comparison_holdout}}
{\begin{tabular}{lllll}\toprule
&Goenen results&With stopping rule \\\cline{3-4}
Dataset&AUC ($\%$)&AUC($\%$)  &  experiments ($\%$)\\\midrule
Nuclear Receptor&82.4&81.7&52.9\\
GPCR&85.7&81.6&39.3\\
Ion Channel&79.9&83.8&44.2\\
Enzyme&83.2&77.8&29.7\\\bottomrule
\end{tabular}}{}
\end{table}

\subsection*{Comparison Of Stopping Rules}

We compared the stopping criteria overall uncertainty (OU) and minimum expected error (MEE) with a fixed threshold as well as an adapted threshold based on label consistency as described \cite{Zhu2010} with the stopping method based on predicted accuracy \cite{Naik2013} in Table \ref{tab:StoppingCriteria}. As in \cite{Zhu2010} we use the absolute difference of the percentage of experiments completed at the stopping time-point to the percentage of experiments completed at the best stopping time (BST) averaged over four data sets ($\Delta_{ave}$) to evaluate different stopping criteria. In \cite{Zhu2010} the BST is defined as the time-point, when the classifier first reaches the highest performance. The predicted accuracy (PA) method with threshold $0.9$ produces the smallest average error to the BST. Both MEE and OU perform two to three times worse than the PA method, even with the adaptive threshold method. The fixed threshold for OU and MEE fails on average, because each of the four data sets has a different optimal threshold for OU and MEE. The maximum uncertainty (MU) and the selected accuracy (SA) stopping criteria \cite{Zhu2010} could not be applied, since those curves are not continuous on these data sets. 
\begin{table}[!t]
\caption{Average difference between the BST point and the stopping point chosen by various stopping rules, over all evaluation data sets.  OU=Overall Average Uncertainty, MEE=Minimum Expected Error PA=Predicted Accuracy. The value in the brackets denotes the threshold. The smaller the difference $\Delta_{ave}$ value is, the better the stopping criterion is. \label{tab:StoppingCriteria}}
{\begin{tabular}{llllll}\toprule
Methods&OU(0.12)& OU(0.09)&OU(0.06)&OU(0.03)&OU(adapted)\\
$\Delta_{ave}$(\%)&40.1 ($\pm$ 12.2)& 33.8 ($\pm$ 17.8)&40.1 ($\pm$ 21.3)&50.9 ($\pm$ 5.4)&28.2 ($\pm$ 29.1)\\

Methods& MEE(0.12)& MEE(0.09)&MEE(0.06)&MEE(0.03)&MEE(adapted)\\
$\Delta_{ave}$(\%)& 40.1 ($\pm$ 11.7)& 38.3 ($\pm$ 12.7)&36.1 ($\pm$ 13.4)&40.6 ($\pm$ 12.1)&30.3 ($\pm$ 12.6)\\

Methods&PA(0.85)& PA(0.9)&PA(0.95)& \\
$\Delta_{ave}$(\%)&32.8 ($\pm$ 8.8)& \textbf{13.7} ($\pm$ 11.3)&22.1 ($\pm$ 15.4)& &\\

\bottomrule
\end{tabular}}{}
\end{table}

\section{Conclusions and Discussion}

We have presented an active learning method for prediction of drug-target interactions based on kernelized matrix factorization. Building on prior work \cite{Goenen2012}, our model can efficiently leverage prior information through kernels to achieve high predictive accuracy. We have furthermore shown that our method can significantly improve the prediction task for drug-target interactions when only a limited number of experiments can be performed. For three real-world data sets with high uniqueness values, the active learning strategy achieves $99\%$ accuracy with 2-3 times fewer experiments than a random sampling strategy. It is important to note that our goal was not to choose the best possible matrix completion method for these specific datasets, but to show that a good method can be used as a basis for active learning to dramatically reduce further experimentation.

Please note, that the presented framework is not limited to KBMF only. Any other model for drug target prediction could be apllied that produces outputs for drug-target scores which can be converted into probabilities. Furthermore the selection strategy uncertainty sampling could be replaced by other active learning strategies (i.e. diversity sampling) to learn new traces on simulated data. The active learning features could be improved by feature selection methods.

For a practitioner to realize these advantages, we have provided a method for estimating the accuracy of an actively learned model using only experimental results already collected; this estimated accuracy is generally a lower bound of the true accuracy of the model. We have shown that this method, calibrated from simulation data, accurately assesses the active learner performance on our real-world data. We have also shown that by applying a stopping rule learned on the simulated data, only half of the experiments are needed to achieve similar accuracies on holdout data. We conclude that active learning driven experimentation is a practical solution to large experimental problems in which time or expense make exhaustive experimentation undesirable.

\end{document}